\documentclass[aps,prx,preprint,letterpaper,groupedaddress,longbibliography]{revtex4-1}
\usepackage[english]{babel}
\usepackage[utf8]{inputenc}
\usepackage{mathtools}
\usepackage{braket}
\usepackage{xcolor}
\usepackage[symbol]{footmisc}
\usepackage{amsmath,amssymb}
\usepackage{caption}
\usepackage{subcaption}
\usepackage{scalerel}
\usepackage{tikz}
\usetikzlibrary{svg.path}
\definecolor{orcidlogocol}{HTML}{A6CE39}
\tikzset{
  orcidlogo/.pic={
    \fill[orcidlogocol] svg{M256,128c0,70.7-57.3,128-128,128C57.3,256,0,198.7,0,128C0,57.3,57.3,0,128,0C198.7,0,256,57.3,256,128z};
    \fill[white] svg{M86.3,186.2H70.9V79.1h15.4v48.4V186.2z}
                 svg{M108.9,79.1h41.6c39.6,0,57,28.3,57,53.6c0,27.5-21.5,53.6-56.8,53.6h-41.8V79.1z M124.3,172.4h24.5c34.9,0,42.9-26.5,42.9-39.7c0-21.5-13.7-39.7-43.7-39.7h-23.7V172.4z}
                 svg{M88.7,56.8c0,5.5-4.5,10.1-10.1,10.1c-5.6,0-10.1-4.6-10.1-10.1c0-5.6,4.5-10.1,10.1-10.1C84.2,46.7,88.7,51.3,88.7,56.8z};}}
\newcommand\orcidicon[1]{\href{https://orcid.org/#1}{\mbox{\scalerel*{
\begin{tikzpicture}[yscale=-1,transform shape]
\pic{orcidlogo};
\end{tikzpicture}
}{|}}}}
\usepackage[colorlinks,citecolor=red,urlcolor=blue,bookmarks=false,hypertexnames=true]{hyperref} 

\newcommand{\kwv}{\mathbf{k}}
\newcommand{\qwv}{\mathbf{q}}
\newcommand{\pwv}{\mathbf{p}}

\newcommand{\fkm}{f_{m\kwv}}
\newcommand{\fk}{f_\kwv}

\newcommand{\perturbo}{\textsc{Perturbo}}
\newcommand{\mobiunits}{cm$^2$V$^{-1}$s$^{-1}$}
\newcommand{\gam}{$\Gamma$}
\newcommand{\kvcm}{kV\ cm$^{-1}$}

\begin{document}

\title{High-field transport and hot electron noise  in GaAs from first principles:  role of two-phonon scattering}

\author{Peishi S. Cheng \orcidicon{0000-0002-3513-9972}}
\author{Jiace Sun
\orcidicon{0000-0002-0566-2084}}
\author{Shi-Ning Sun \orcidicon{0000-0002-5984-780X}}
\author{Alexander Y. Choi \orcidicon{0000-0003-2006-168X}}
\author{Austin J. Minnich \orcidicon{0000-0002-9671-9540}}
\email{aminnich@caltech.edu}

\affiliation{Division of Engineering and Applied Science, California Institute of Technology, Pasadena, CA, USA}

\date{\today}

\begin{abstract}

High-field charge transport  in semiconductors is of fundamental interest and practical importance. While the \textit{ab initio} treatment of low-field transport is well-developed, the treatment of high-field transport is much less so, particularly for multi-phonon processes that are reported to be relevant in GaAs. Here, we report a calculation of the high-field transport properties and current power spectral density (PSD) of hot electrons in GaAs from first principles including on-shell two-phonon (2ph) scattering. The on-shell 2ph scattering rates are found to qualitatively alter the high-field distribution function by increasing both the momentum and energy relaxation rates as well as contributing markedly to intervalley scattering. This finding reconciles a long-standing discrepancy regarding the strength of intervalley scattering in GaAs as inferred from transport and optical studies. The characteristic non-monotonic trend of PSD with electric field is not predicted at this level of theory. Our work shows how \textit{ab initio} calculations of high-field transport and noise may be used as a  stringent test of the  electron-phonon interaction in semiconductors.

\end{abstract}

% , indicating that off-shell 2ph or other processes play a fundamental role in high-field transport

\maketitle
\section{Introduction}\label{sec:Introduction}

High-field charge transport and fluctuation phenomena in semiconductors are of fundamental and practical interest for semiconductor devices \cite{sze2007, Bareikis_book, Hartnagel_2001}. Early work on high-field charge transport  in semiconductors focused on dielectric breakdown in polar semiconductors, establishing an early treatment of electronic interactions with polar longitudinal optical phonons \cite{frohlich1937}. Subsequent studies examined the nonlinear variation of drift velocity with electric field in elemental semiconductors \cite{ryder1951,ryder1953, shockley1951}, current instabilities in III-V semiconductors known as the Gunn effect \cite{gunn_1963, Ridley_1963},  experimental characterization of the negative differential resistance region associated with the Gunn effect \cite{ruchkino1967,braslau1967,acket1967}, and measurement of the warm electron and sub-millimeter wave mobility that provided insight into the energy relaxation time \cite{vlaardingerbroek1968, kuchar1972, Abe_1972, glover1973}. Initial theoretical studies employed model distribution functions \cite{conwell1966_eloss, Butcher_1965} or numerically solved the Boltzmann equation under various approximations \cite{conwell1966, conwell_1968, rees1969freq} to investigate high-field transport phenomena such as energy relaxation and intervalley scattering processes. Beginning in the 1970s, Monte Carlo calculations became the predominant method for modeling high-field transport \cite{fawcett1970, ruch1970, littlejohn1977}, enabling the simulation of various phenomena  across a range of electric fields, temperatures, and geometries \cite{reggiani_1983}.

In addition to measurement of typical transport properties, velocity fluctuations in semiconductors were characterized using measurements of the current power spectral density (PSD). Owing to the fluctuation-dissipation relation, close to equilibrium the PSD does not provide additional information about charge transport beyond that contained in the mobility \cite{Kogan_1996, Callen_1952, Nyquist_1928}. However, away from equilibrium, the PSD contains qualitatively new information because it characterizes the fluctuations about a non-equilibrium steady-state distribution, in contrast to observables that characterize the mean of the steady-state distribution like drift velocity \cite{kleinert2010, Hartnagel_2001}. The first reported measurement of velocity fluctuations in a semiconductor was of the transverse noise temperature in Ge \cite{erlbach1962}. PSD was later obtained by measuring the diffusion coefficient using the relation between PSD and diffusion coefficient at sufficiently low frequencies \cite{Price_1965}. Initial measurements of the PSD versus electric field in GaAs exhibited a peak at an electric field around the onset of negative differential resistance ($\sim 2-3$ \kvcm) \cite{ruchkino1968}. Although the quantitative results may have been complicated by a non-negligible time-domain response of the electric circuit used in the experiment \cite{glisson1980}, later measurements using microwave pulses  confirmed the trend \cite{Bareikis_1980, gasquet1985}. The first numerical investigations of noise phenomena in GaAs focused primarily on calculating the diffusion coefficient at high fields \cite{fawcett_1969}, noting that the high-field diffusion coefficient differed substantially from that predicted using the Einstein relation \cite{jacoboni1974, quaranta1973}. Later, Monte Carlo calculations \cite{pozela1978, pozela1980, Bareikis_1980} and analytical models \cite{Xing_1988_2, Stanton_1987_1, Stanton_1987_2} of transport in GaAs were used to study intervalley processes and their role in the Gunn effect and PSD peak. Subsequent works focused on the effect of short channels \cite{Bareikis_1986} or impurities \cite{bareikis1989} on PSD at cryogenic temperatures, Monte Carlo modeling of noise in modern heterostructure devices \cite{Mateos_2000, Mateos_2015}, and analytical models of PSD in graphene \cite{rustagi2014}.

Intervalley scattering in GaAs has been the subject of substantial experimental and theoretical study due to its role in producing negative differential resistance  \cite{Jacoboni2010} and non-monotonic features of the PSD versus electric field \cite{Shockley_1966, Price_1960, Bareikis_1994, nougier1994}. Early theoretical works derived symmetry selection rules for intervalley scattering, concluding that only LA and LO phonons could mediate intervalley coupling between states at the $\Gamma$ and L points \cite{birman1966}. Diverse experimental and numerical methods have reached conflicting conclusions regarding the strength of intervalley scattering in GaAs as quantified by the intervalley deformation potential (IDP), $D_{\Gamma\mathrm{L}}$ \cite{mickevicius_1990}.  Transport studies involving measurements of  PSD  \cite{pozela1978,pozela1980}, I-V curves in sub-micron structures \cite{Gruzinskis_1988}, and threshold field versus stress \cite{Adams1977} interpreted using Monte Carlo simulations with semi-empirical scattering rates \cite{mitskyavichyus1986} concluded that the intervalley scattering strength must be weak ($D_{\Gamma\mathrm{L}} \sim 2\times10^8$ eV cm$^{-1}$) to match trends of experimental data. On the other hand, experiments based on photoluminescence of optically excited carriers inferred a markedly larger value ($D_{\Gamma\mathrm{L}} \sim 8\times10^8$ eV cm$^{-1}$) \cite{shah1987, ulbrich1989, kash1993, karlik1987, mirlin1987}. Recent first-principles calculations \cite{Sjakste2007} support the larger value of the deformation potential due to contributions from non-longitudinal phonons, as selection rules are relaxed away from high symmetry points \cite{zollner1990_JAP}. Advances in experimental methods have enabled the relaxation of photoexcited electrons to be monitored with momentum and energy resolution, providing insights into the effect of intervalley scattering on the differing timescales of momentum and energy relaxation \cite{sjakste2018review, sjakste2018, tanimura2016}. Despite these experimental advances, the discrepancy in the intervalley scattering strength inferred from various experiments remains unresolved.

\textit{Ab initio} methods may aid in resolving such discrepancies by providing a parameter-free treatment of the electron-phonon (e-ph) interaction and charge transport processes \cite{Giustino_2017, Bernardi_2016}. The low-field mobility has been computed for diverse materials \cite{ponce_2020} including Si \cite{restrepo2009, Ponce_2018}, GaN \cite{ponce2019gan}, GaAs \cite{Zhou_2016}, hybrid perovskites \cite{Motta2015, Zhao2016}, two-dimensional materials \cite{borysenko2010, kaasbjerg2012, 2dmobi2020}, and other compound semiconductors \cite{ponce2021}. Advances to \textit{ab initio} e-ph theory continue to be reported, with examples such as the $GW$ corrections to the e-ph interaction \cite{Ponce_2018,gwpt_2019} and inclusion of the quadrupole interaction \cite{brunin2020,quadprb2020}. Within the Boltzmann transport framework, works have studied phonon drag  \cite{protik2020} and magnetotransport \cite{dhruv2021}. For GaAs, achieving quantitative accuracy in the first-principles calculation of low-field mobility remains a subject of ongoing work, with discrepancies between initial \textit{ab initio} calculations \cite{Zhou_2016,Liu_2017} largely ascribed to differences in the band structure with different effective masses and valley separations \cite{Ma_2018}. A recent calculation of electron mobility including higher-order terms in which electrons are sequentially scattered by two phonons has indicated that these processes are non-negligible  \cite{lee2020}. For high-field transport, only recently has the \textit{ab initio} framework been applied, with drift velocity curves calculated by explicitly time-stepping the Boltzmann equation to steady state \cite{maliyov2021} or in combination with Monte Carlo simulations \cite{ghosh2017}. The \textit{ab initio} treatment of electronic noise at high fields is comparatively lacking, with a first-principles framework for electronic noise only recently reported but restricted to the warm electron regime \cite{warmelectrons}. Finally, the present level of theory for either transport or PSD has not been tested in the hot electron regime in polar semiconductors in which energy relaxation and intervalley scattering are of fundamental importance. \textit{Ab initio} calculations of high-field transport and hot electron noise offer a new test of the accepted theory of e-ph interactions in semiconductors by probing fluctuations about a non-equilibrium steady-state distribution function.

Here, we report an \textit{ab initio} calculation of the drift velocity and PSD of hot electrons in GaAs for electric fields up to 5 \kvcm~ including on-shell 2ph scattering. We find that this higher-order e-ph scattering process plays a fundamental role in all aspects of high-field transport, particularly by increasing the average energy dissipated to the lattice per scattering event and increasing the intervalley scattering rate. This observation  provides an explanation for the incompatible values of IDP as inferred from transport and optical studies.  We find that the characteristic non-monotonic trend of PSD with electric field is not predicted even with on-shell 2ph scattering.  This finding highlights the use of the PSD as a rigorous test of the theory of e-ph interactions in semiconductors.

% The discrepancy may arise from  higher-order electron-phonon interactions or experimental non-idealities.
% , and it is a topic of future study.

%  which occurs around the region of negative differential resistance

% Our work also demonstrates that \textit{ab initio} calculations of high-field transport and PSD can be used as a rigorous test of the theory of e-ph interactions in semiconductors.

% , suggesting that off-shell 2ph processes are  of fundamental importance for high-field transport

%%% THEORY
\section{Theory}\label{sec:Theory}

\subsection{Overview}

The details of the \textit{ab initio} formalism to compute transport and noise coefficients beyond the cold electron regime are given in Ref.~\cite{warmelectrons}. Here, we summarize the results and indicate the necessary changes to extend the method to the hot electron regime.

The Boltzmann transport equation (BTE) for spatially homogeneous electrons subjected to an electric field $\mathbf{E}$ is:
 \begin{equation}\label{BTE}
    \frac{\partial f_{m\kwv}}{\partial t} + \frac{e\mathbf{E}}{\hbar}\cdot \nabla_{\kwv}\fkm = \mathcal{I}[\fkm]
 \end{equation}
Here, $f_{m\kwv}$ is occupation of the electron state at wave vector $\mathbf{k}$ with band index $m$, and $e$ is the fundamental charge. $\mathcal{I}$ is the collision integral that describes scattering of electrons by phonons, the dominant form of scattering near room temperature for non-degenerate carrier concentrations \cite{Bernardi_2016}. At steady state, the time derivative term vanishes by definition and the solution of the Boltzmann equation is the non-equilibrium steady-state distribution. Our work focuses on the conduction band of GaAs for which there are no interband transitions in the energies of interest, so we omit the electron band indices in the remaining equations for simplicity.

\subsection{Collision integral at high fields}

The 1ph e-ph collision integral is given by: 
\begin{equation}\label{collint}
     \mathcal{I}[\fk] = -\frac{2 \pi}{\hbar} \frac{1}{N} \sum_{\nu \qwv}  \left|g_{\kwv}^{\kwv + \qwv}\right|^2 
     \bigg(\delta(\epsilon_{\kwv}-\hbar \omega_{\nu \qwv}-\epsilon_{\kwv+\qwv})H^{\mathrm{em}}_\kwv + \delta(\epsilon_{\qwv} + \hbar \omega_{\nu\qwv} -\epsilon_{\kwv+\qwv}) H^{\mathrm{abs}}_\kwv \bigg)
\end{equation}
where the sum is over phonon wave vector $\mathbf{q}$ and polarization $\nu$ for scattering phonons which satisfy momentum conservation, the delta functions ensure energy conservation, and $g_{\kwv}^{\kwv + \qwv}$ is the e-ph matrix element. $N$ is the total number of $\mathbf{q}$-points sampled from the Brillouin zone. $H^{\mathrm{em}}_\kwv$ and $H^{\mathrm{abs}}_\kwv$ weight the scattering probabilities to account for the electron and phonon occupations. In general, the weights are nonlinear functions of the electron occupations (see Eqs.~(11.121) and (11.125) in Ref.~\cite{Mahan_2011}).

The non-linear character of the collision integral makes the numerical solution of the BTE challenging at high electric fields.  In our previous framework for warm electrons \cite{warmelectrons}, the collision integral took the typical form in the literature in which the weights are linearized about a deviational occupation $\Delta f_\kwv$ as defined by $f_\kwv = f^0_\kwv + \Delta f_\kwv$, where $f^0_\kwv$ is the equilibrium Fermi-Dirac distribution. The weights associated with this linearization are:
\begin{eqnarray}\label{lowfieldweights}
    H^{\mathrm{em}}_\kwv &= \Delta f_\kwv(N_\qwv + 1 - f^0_{\kwv+\qwv}) - \Delta f_{\kwv+\qwv}(N_\qwv + f_\kwv^0) \\
    H^{\mathrm{abs}}_\kwv &= \Delta f_\kwv(N_\qwv + f^0_{\kwv+\qwv}) - \Delta f_{\kwv+\qwv}(N_\qwv + 1 - f^0_{\kwv})
\end{eqnarray}
These weights are obtained by assuming $\Delta f_{\mathbf{k}} \ll f_{\mathbf{k}}^0$ and neglecting higher order terms of the form $\Delta f_{\mathbf{k}} \Delta f_{\kwv + \qwv}$. This assumption is violated at high electric fields as the deviational occupations far exceed the equilibrium occupations for high-energy electron states. However, for non-degenerate electrons, the distribution function values are always much less than the phonon occupations ($f_{\mathbf{k}} \ll N_{\mathbf{q}}$) even at high fields. Therefore, all terms involving the equilibrium electronic occupations can be neglected. The resulting weights are:

\begin{eqnarray}\label{highfieldweights}
    H^{\mathrm{em}}_\kwv &= \Delta f_\kwv(N_\qwv + 1) - \Delta f_{\kwv+\qwv} N_\qwv \\
    H^{\mathrm{abs}}_\kwv &= \Delta f_\kwv N_\qwv - \Delta f_{\kwv+\qwv}(N_\qwv + 1)
\end{eqnarray}

With these weights, the BTE again corresponds to a linear system of equations that can be solved to obtain the drift velocity and mobility in the hot electron regime. Since we solve the linear system with non-diagonal elements in the collision integral, we obtain the full solution to the BTE beyond the relaxation time approximation. With the collision integral modified for high electric fields, the remainder of the steady-state calculation follows the same framework as given in Ref.~\cite{warmelectrons}. In particular, the drift velocity is obtained using Eq.~20 of Ref.~\cite{warmelectrons}.  In this work, we restrict $E \leq 5$ \kvcm, where $E$ is the magnitude of the electric field, due to the increasing computational cost of incorporating states at higher energies that become occupied as the electric field increases (see Sec.~\ref{methods} for further discussion).

The physical meaning of neglecting the electron occupations in the weights is that the e-ph scattering rate is independent of the electron occupations at the initial and final states of the scattering process for non-degenerate carrier concentrations. The approximation is particularly well-satisfied for GaAs owing to its relatively low Debye temperature of 360 K \cite{LandoltBornstein2002}, yielding phonon occupations $N_q \sim 1$ even at optical phonon energies. Additionally, at higher carrier concentrations, scattering processes besides  e-ph scattering such as electron-electron interactions may become non-negligible \cite{bernardipnas_2015, Collet1993}. In this case, an additional correlation between electron occupations arises that complicates the calculation of the current PSD \cite{GGK_1979}. We do not consider such correlation in the present computational framework.

\subsection{\textit{Ab initio} computation of PSD}

We briefly review the formalism to compute the current PSD once the steady-state distribution function is obtained. The PSD is given by the Fourier transform of the autocorrelation of current fluctuations (Eq.~17 of Ref.~\cite{warmelectrons}). The governing equation for the autocorrelation function is again the Boltzmann equation as in the steady case \cite{GGK_1969}. After some manipulation (given in Sec.~II.B of Ref.~\cite{warmelectrons}), the PSD can be obtained by solving the following equation:

\begin{equation}
\label{effdist}
    \left[i\omega\mathbb{I} + \Lambda\right]g_{\kwv} = f_{\mathbf{k}}^s (v_{\kwv,\alpha} - V_{\alpha})
\end{equation}
where $\Lambda$ is the relaxation operator that contains the electric field term and the e-ph collision integral; $\omega$ and $\alpha$ are the frequency and Cartesian direction, respectively, for which the noise is calculated; $f_{\kwv}^s$ is the steady-state occupation for the state at wave vector $\kwv$ for the given field; $v_{\kwv,\alpha}$ is the group velocity for state $\kwv$; and $V_\alpha$ is the drift velocity.

The solution to this second Boltzmann equation, $g_{\kwv}$, has been denoted the ``effective distribution function" \cite{rustagi2014}. The physical meaning of the right-hand side of Eq.~\ref{effdist} is that the PSD is larger for steady-state distributions with occupation in states for which there is a larger difference between the group velocity and the drift velocity, roughly corresponding to distributions with larger variance. Finally, the longitudinal current PSD is obtained by a Brillouin zone integration of the effective distribution function given by the expression below:

\begin{equation}
\label{noisesum}
    S_{j_{\alpha}j_{\alpha}}(\omega) =
    2 \bigg(\frac{2 e}{\mathcal{V}_0}\bigg)^2 \Re \left[ \sum_{\mathbf{k}} v_{\kwv,\alpha} g_{\kwv} \right]
\end{equation}
where $S_{j_{\alpha}j_{\alpha}}(\omega)$ is the current PSD at frequency $\omega$, with electric field in direction $\alpha$. The factor in front of the sum contains the fundamental charge $e$, and the volume of the supercell $\mathcal{V}_0$. 

\subsection{Two-phonon (2ph) scattering}\label{2phtheory}
The computational framework for the first-principles calculation of 2ph scattering, where electrons are scattered by two consecutive one-phonon events, was recently developed and reported to be non-negligible for low-field mobility in GaAs \cite{lee2020}. As we will show, the level of theory with first-order e-ph scattering where electrons are scattered by one phonon (1ph) is insufficient for high-field transport and hot electron noise. Therefore, we included scattering from 2ph processes with approximations to ensure computational tractability for the high-field case. 

According to  Ref.~\cite{lee2020}, for $\kwv \neq \kwv'$ the 2ph collision matrix may be written as:

\begin{equation}
\label{eq:2ph_integral}
    \Theta^{\mathrm{(2ph)}}_{\kwv,\kwv'} = - \frac{2\pi}{\hbar}\frac{1}{N^2}\sum_{\qwv + \pwv = \kwv-\kwv'}\sum_{\nu\mu} \left[\tilde{\Theta}^{\mathrm{(1e1a)}} + \tilde{\Theta}^{\mathrm{(2e)}} + \tilde{\Theta}^{\mathrm{(2a)}} \right]
\end{equation}

where $N$ is the number of phonon points sampled from the Brillouin zone, and sums are over all pairs of phonons that couple two electronic states, with the second phonon identified by branch index $\mu$ and wave vector $\pwv$. The minus sign is to conform to the sign convention of Ref.~\cite{warmelectrons}. The diagonal element of the scattering matrix gives the scattering rate of the state and is given by: 

\begin{equation}
\label{eq:2ph_rates}
    \Theta_{\kwv', \kwv'}^{\mathrm{(2ph)}} = \Gamma^{\mathrm{(2ph)}}_{\kwv'} = - \sum_{\kwv\neq \kwv'} \Theta^{\mathrm{(2ph)}}_{\kwv,\kwv'}
\end{equation}

The superscripts in Eq.~\ref{eq:2ph_integral} refer to the three types of 2ph processes: phonon emission combined with phonon absorption (1e1a), emission of two phonons (2e), and absorption of two phonons (2a). The contribution of each type of 2ph process, indexed by superscript $i$, is given by:

\begin{equation}
    \tilde{\Theta}^{(i)} = A^{(i)}  W^{(i)} \delta 
    (\epsilon_\kwv - \epsilon_{\kwv'} - \alpha^{(i)}_{\pwv}\omega_{\mu\pwv} - \alpha^{(i)}_{\qwv}\omega_{\nu\qwv})
\end{equation}
where $A^{(i)}$ is the weighting factor based on phonon and electron occupations (Eq.~4 in Ref.~\cite{lee2020}), $W^{(i)}$ is the 2ph process amplitude, and the constants $\alpha^{(i)}$ are determined by the type of scattering process, taking on the values:
\begin{table}[h]
\centering
\begin{tabular}{ccc}
    $\alpha_\pwv^{\mathrm{(1e1a)}} = 1$, & 
    $\alpha_\pwv^{\mathrm{(2e)}} = 1$, & 
    $\alpha_\pwv^{\mathrm{(2a)}} = -1$, \\
    $\alpha_\qwv^{\mathrm{(1e1a)}} = -1$, &
    $\alpha_\qwv^{\mathrm{(2e)}} = 1$, & 
    $\alpha_\qwv^{\mathrm{(2a)}} = -1$\\
\end{tabular}
\end{table}

The 2ph process amplitude is given by:
\begin{equation}
\label{2pheq}
\begin{aligned}
    W^{(i)} %&= |W^{(i)}(\qwv, \pwv, \alpha_\pwv) + W^{(i)}(\pwv, \qwv, \alpha_\qwv)|^2
    %\\
    &= \left | \left(
    \frac{ g_{\nu}(\kwv,\qwv)g_{\mu}(\kwv+\qwv,\pwv) }
         { \epsilon_{\kwv'}-\epsilon_{\kwv+\qwv}+\alpha^{(i)}_\pwv\omega_{\nu\pwv} + i\eta - \Sigma_{\kwv+\qwv} } 
    +\frac{ g_{\mu}(\kwv,\pwv)g_{\nu}(\kwv+\pwv,\qwv) }
         { \epsilon_{\kwv'}-\epsilon_{\kwv+\pwv}+\alpha^{(i)}_\qwv\omega_{\nu\qwv} + i\eta - \Sigma_{\kwv+\pwv} }
    \right)\right |^2
\end{aligned}
\end{equation}
where $g_{\nu}(\kwv,\qwv)$ is the one-phonon matrix element corresponding to coupling between an electron at $\kwv$ scattering to an electron at $\kwv+\qwv$ through a phonon of mode $\nu$ at $\qwv$, and so on for the other matrix elements. The $\epsilon$ correspond to the band eigenvalues, $\omega$ are the phonon energies, $i\eta$ is an infinitesimal required to prevent divergences in the denominator, and $\Sigma_{\kwv+\qwv}$ is the self-energy of the electron at $\kwv+\qwv$.

The 2ph framework differs from the 1ph framework in several ways; we note two particularly important differences indicated by Eq.~\ref{2pheq}. First, the intermediate electron state reached after being scattering by the first phonon ($\epsilon_\kwv \pm \hbar\omega_\qwv$) is a virtual state that does not necessarily have the band energy ($\epsilon_{\kwv+\qwv}$) at the corresponding point in the Brillouin zone. If the virtual state energy coincides (does not coincide) with the band energy, the 2ph process is called ``on-shell'' or ``resonant'' (``off-shell'' or ``non-resonant''). The difference between the virtual energy and the band energy is called the off-shell extent. Second, the amplitude of a given 2ph scattering process depends on the self-energy of the intermediate state ($\Sigma$ in the denominators of Eq.~\ref{2pheq}). We consider only the imaginary part of the self-energy which is directly proportional to the scattering rate at the intermediate state. Since the scattering rate should include both 1ph and 2ph scattering, the calculation for 2ph processes requires iteration until self-consistency, where the initial iteration approximates the intermediate state scattering rate as containing only 1ph processes, and the resulting 2ph scattering rate is used in the next iteration. 

Including 2ph scattering at the fully \textit{ab initio} level for high-field transport is computationally prohibitive. We employ several approximations to make the calculation feasible.  First, we found empirically that increasing the number of self-consistent iterations beyond three did not lead to qualitative changes in the trend of computed observables, and we therefore employed three iterations rather than ten as in Ref.~\cite{lee2020}. Second, consistent with our approximations used to extend the 1ph collision integral to high fields, we also neglect the electron occupation terms in the 2ph weights of $A^{(i)}$, given in Eq.~4 of Ref.~\cite{lee2020}. 

Finally, we restrict the 2ph processes to include only those with  an intermediate state that is within a specified threshold of off-shell extent, meaning that the process is nearly on-shell.  This on-shell approximation is expected to capture many relevant 2ph processes owing to the 2ph rate being inversely proportional to the square of the off-shell extent (c.f. the denominator of Eq.~\ref{2pheq}). Although these on-shell 2ph processes consist of successive 1ph pathways, they directly couple electronic states that are not directly coupled by 1ph processes and may therefore qualitatively alter the momentum and energy relaxation compared to the 1ph level of theory. 

The on-shell restriction leads to subtle differences with the formulation presented in Ref.~\cite{lee2020}. Considering the two terms on the right-hand side of Eq.~\ref{2pheq}, if one of the terms is on-shell, the other term is likely to be off-shell. The term that is off-shell is therefore neglected. However, the neglect of this term means that the factor of 1/2 accounting for double counting in 2e and 2a processes (Eq.~4 of Ref.~\cite{lee2020}) should not be included in the present formulation. In addition, while both 1a1e and 1e1a processes are automatically included when using the full expression of Eq.~\ref{2pheq} for 1e1a processes, with the off-shell term neglected both 1a1e and 1e1a processes must be explicitly incorporated.  The 2ph rates we obtain with these approximations we term on-shell 2ph for the remainder of the paper. Properties computed using both 1ph and on-shell 2ph scattering rates are denoted 1+2ph.

\section{Numerical methods}\label{methods}
The calculation of high-field drift velocity and PSD takes the electronic structure and e-ph matrix elements as inputs, which are computed for GaAs from first principles using Density Functional Theory (DFT) and Density Functional Perturbation Theory (DFPT) with \textsc{Quantum Espresso} (QE) \cite{giannozzi_qe_2009, giannozzi_qe_2017}. Following Ref.~\cite{warmelectrons}, the calculation uses an $8\times 8\times 8$ coarse grid, a plane wave cutoff of 72 Ryd, a lattice parameter of 5.556 $\textrm{\AA}$, and a non-degenerate carrier concentration of 10$^{15}$ cm$^{-3}$.

We implement the PSD calculation with additional routines that take as input the e-ph matrix elements obtained from \perturbo~\cite{perturbo_2021}, which performs the Wannier interpolation of the QE data to the fine grids necessary for transport calculations. We use a fine grid of $250\times 250\times 250$ for all calculations in the 1ph framework. Once the interpolated e-ph matrix elements are obtained using unmodified \perturbo\ routines, we explicitly construct the high-field collision matrix given by Eqs.~\ref{collint} and \ref{highfieldweights}. The delta functions in Eq.~\ref{collint} are approximated by a Gaussian with a smearing parameter of 5 meV. As in Ref.~\cite{warmelectrons}, the drift term (second term in the left hand side of Eq.~\ref{BTE}) is implemented using the finite-difference scheme of Refs. \cite{mostofi_2008,Pizzi_2020}. We solve the resulting linear systems using a Fortran implementation of the Generalized Minimal Residual (GMRES) method \cite{gmres_2005}. We note that the recent work of Ref.~\cite{maliyov2021} obtained the steady-state distribution by explicitly time-stepping the BTE to steady state; here, we solve for the steady-state distribution directly using numerical linear algebra.

In \perturbo, an energy window is specified to limit the points sampled in the Brillouin zone to only those in the relevant energy range. As the electric field increases, electrons are driven to higher energies compared to the energies relevant to thermal equilibrium at room temperature, so this energy window must be larger than the window used for low-field mobility calculations. Increasing the energy window significantly increases the number of points in part due to the inclusion of the L valley. The primary limitation on computational tractability is the number of k-points sampled since the size of the collision matrix grows quadratically with the number of k-points. We find that for the electric fields of interest ($\leq 5$ \kvcm), an energy window of 360 meV above the conduction band minimum is sufficient, yielding around 52,000 k-points. We found that increasing the energy window to 400 meV had no qualitative effect on the calculated trends of drift velocity and PSD versus electric field.

For the 2ph calculations, we include only processes with an off-shell extent of 25 meV or less, meaning the intermediate virtual state is within 25 meV of the band energy. We find that increasing this tolerance to 30 meV increases the 2ph rates by only 1.2\%. When calculating noise and transport quantities with 2ph scattering, we are limited by computational tractability to a fine grid of $200\times 200\times 200$.

% and a energy window of 360 meV above the conduction band minimum.

\section{Results}

\subsection{Drift velocity with 1ph scattering}

\begin{figure}[h]
\centering
\includegraphics[width=0.6\textwidth]{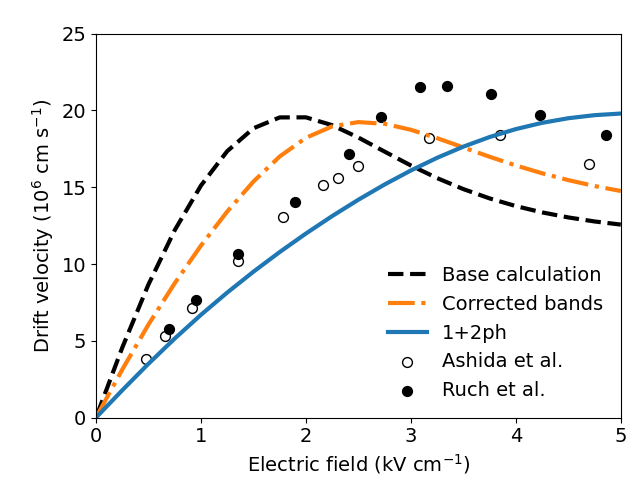}
\caption{Drift velocity versus electric field for the base calculation (dashed black curve), the corrected-bands case (dot dashed orange curve), and the 1+2ph case (solid blue curve), as described in the text. The inclusion of on-shell 2ph processes gives the best agreement of the three cases with the experimental drift velocity measurements of Ruch et al. \cite{ruchkino1967} and Ashida et al. \cite{ashida1974}.}
\label{fig:hfobserv}
\end{figure}

We first consider the drift velocity versus electric field at the 1ph level of theory,  presented in Fig.~\ref{fig:hfobserv}.  The base calculation is observed to qualitatively reproduce several trends, including the linear increase of the drift velocity with electric field below 1 \kvcm,  followed by a rapid decrease and a region of negative differential mobility. However, consistent with prior reports \cite{lee2020}, the low-field mobility, corresponding to the slope of the drift velocity at low electric fields, is considerably overestimated, with the predicted mobility of 18,570 \mobiunits~exceeding the experimental value of around 8,000 \mobiunits \cite{blakemore1982} by over a factor of two.

Some of the discrepancies in the drift velocity curve can be attributed to inaccuracies in the DFT band structure.  First, the computed effective mass is overestimated compared to experiment (0.055$m_e$  versus 0.067$m_e$, respectively) \cite{Ma_2018}. Second, the minima of the L valley in the DFT band structure are at 250 meV above the conduction band minimum (CBM) instead of 300 meV as in experiments \cite{blakemore1982, aspnes1976}. To quantify the correction to the drift velocity due to the band structure, we replace the energy eigenvalues of states in the $\Gamma$ valley with those calculated using a spherically symmetric band structure model \cite{conwell_1968} with the experimental effective mass of 0.067$m_e$ and a non-parabolicity of 0.64 \cite{Lundstrom_2000}. We also rigidly shift the DFT band energies in the L valleys by 50 meV to achieve the experimental $\Gamma$-L valley separation of 300 meV. We note that while other works have obtained band structures closer to experiment using $GW$ corrections \cite{Ma_2018,Liu_2017,lee2020}, prior analysis for GaAs has argued that the main effect of these corrections is to alter the effective mass rather than the e-ph coupling strength \cite{Ma_2018}. 

The drift velocity versus electric field with these corrections, denoted  ``corrected bands,'' is plotted in Fig.~\ref{fig:hfobserv}. The agreement of the velocity field curves with experiments is improved, with a low-field mobility of 12,674~\mobiunits, but it remains overpredicted.

\subsection{On-shell 2ph scattering}

\begin{figure}[h]
\centering{
\phantomsubcaption\label{fig:2phrates-a}
\phantomsubcaption\label{fig:2phrates-b}
\phantomsubcaption\label{fig:2phrates-c}
\phantomsubcaption\label{fig:2phrates-d}
\includegraphics[width=0.8\textwidth]{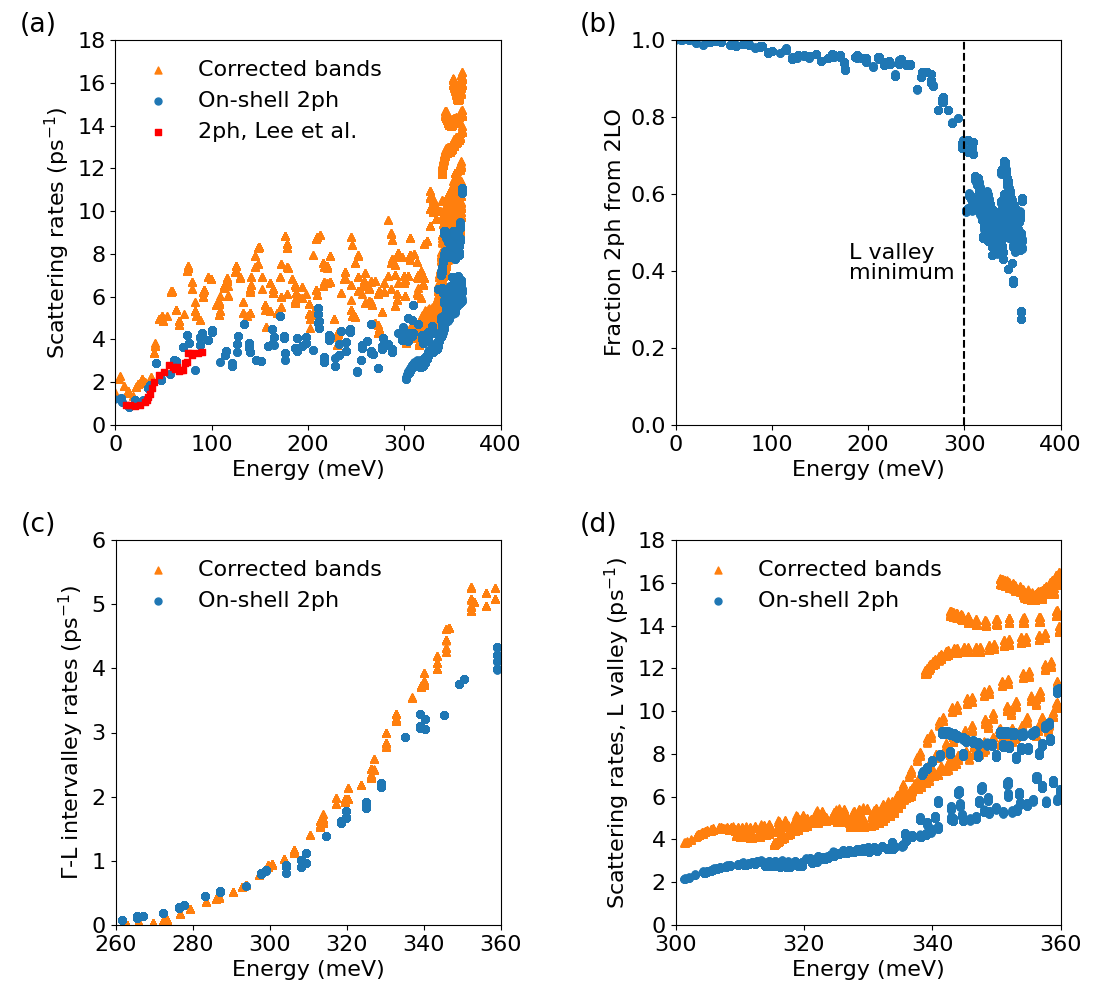}}
\caption{Scattering rates from on-shell 2ph processes versus energy.  (a) 1ph scattering rates from the corrected-bands case (orange triangles), the on-shell 2ph rates (blue dots), and the full 2ph rates iterated to self-consistency as given in Fig.~4 of Ref.~\cite{lee2020} (red squares).  The increase in scattering rates around 0.3 eV is due to the onset of intervalley  processes. (b) Fraction of the on-shell 2ph rates coming from processes mediated by LO phonons only. As intervalley scattering becomes permitted, non-LO phonons participate in 2ph scattering. (c) Intervalley scattering rate for states in \gam~scattered to the L valley. The on-shell 2ph intervalley scattering rate is comparable to that for 1ph  scattering. (d) On-shell 2ph scattering rates in the L valley.}
\label{fig:2phrates}
\end{figure}

It has been reported that the low-field mobility of GaAs, and thus the slope of the drift velocity versus field curve, is overestimated even with corrections to the band structure, and that additional scattering  from 2ph processes is necessary to achieve improved agreement with experiments \cite{lee2020}. To assess the impact of these corrections on the high-field drift velocity, we computed the on-shell 2ph scattering rates as specified in Sections \ref{2phtheory} and \ref{methods}. These calculations employed the corrected band structure described above to facilitate comparison of the effects of 2ph scattering relative to the effective mass correction. 

The effect of the additional on-shell 2ph scattering on the drift velocity is shown in Fig.~\ref{fig:hfobserv}. Consistent with Ref.~\cite{lee2020}, the agreement of the  low-field mobility is improved, with the computed value of 7153 \mobiunits~about 10\% lower than the accepted value of $\sim 8000$ \mobiunits \cite{blakemore1982}. However, the high-field drift velocity with on-shell 2ph scattering, which have not been previously reported, is now underpredicted compared to experiment. We observe that the threshold field for the onset of negative differential resistance is around 5 \kvcm, exceeding the experimental value of around 3.5 \kvcm~\cite{blakemore1982}. 

To gain more insight, we examine various features of the on-shell 2ph rates, which have not yet been reported for energies above 100 meV where processes relevant to high-field transport such as intervalley scattering and energy relaxation occur. In Fig.~\ref{fig:2phrates-a}, we show the on-shell 2ph scattering rates versus energy above the CBM along with the 1ph scattering rates with the corrected bands. We find that the on-shell 2ph rates are comparable to the 1ph rates over the entire energy range up to 360 meV, with a magnitude around half of the 1ph rates.  Between 100 meV and 300 meV, the 2ph scattering rates are roughly constant, and increase above 300 meV. This trend is also observed in the 1ph rates and is attributed to the onset of intervalley scattering. The on-shell 2ph rates are in good agreement with the full 2ph rates reported in Ref.~\cite{lee2020}. This observation suggests that off-shell processes make a negligible contribution to scattering, and that the on-shell 2ph scattering in the present work has largely captured the relevant 2ph processes.

% underestimate is expected given the neglect of most off-shell 2ph processes in our calculation. The full 2ph rates include scattering from virtual states that can be arbitrarily far from the band energy and thus include many additional processes. Although the contribution to the scattering rate of an individual off-shell process decreases with off-shell extent per Eq.~\ref{2pheq}, the large phase space for such processes may compensate so that the final off-shell scattering rate is non-negligible. 

Figure \ref{fig:2phrates-b} shows the fraction of processes involving two LO phonons versus energy. Below 200 meV, greater than 90\% of the on-shell 2ph processes involve only LO phonons, but at higher energies near and above the minimum of the L valleys, a substantial fraction of the 2ph processes involve phonons other than the LO mode. In Ref.~\cite{lee2020}, only 2ph processes involving LO phonons were considered, and the figure shows that such an approximation is justified for the low-energy scattering rates relevant for the low-field mobility. However, it is known that intervalley scattering is mediated through all phonon modes \cite{Liu_2017, Ma_2018, Zhou_2016, Sjakste2007, zollner1990_JAP}, not just the LO mode, thereby explaining why the contribution of non-LO phonons becomes increasingly important at higher energies. For energies exceeding $\sim 320$ meV, around half of the scattering processes involve non-LO phonons.

To assess the magnitude of intervalley scattering due to 2ph processes, we plot the intervalley scattering rates for transitions from the \gam~valley to the L valley in Fig.~\ref{fig:2phrates-c}. The 2ph intervalley rates are of comparable magnitude to those of the 1ph framework. Prior studies of intervalley scattering have not considered the contribution of 2ph processes, which are now seen to be as important as 1ph processes. Figure \ref{fig:2phrates-d} shows the 2ph scattering rates in the L valleys. Here, the scattering rates are around  half of those for 1ph scattering, particularly near the L valley minimum.

% At higher energies, the 2ph rates are a smaller fraction of the 1ph rates due to the dependence of the 2ph rates on the self-energy of the intermediate state in the denominator of the 2ph process amplitude (Eq.~\ref{2pheq}). The scattering rates and thus imaginary part of the self-energies for the intermediate states become larger at higher energies, and the 2ph process amplitude decreases, reducing the strength of 2ph scattering.

% Prior studies of intervalley scattering have not included 2ph processes, which affects the interpretation  that affect charge transport properties as we discuss in Section \ref{missingpeak}.

\subsection{Valley occupation and high field distribution}

\begin{figure}[h]
\centering{
\phantomsubcaption\label{fig:hfdist-a}
\phantomsubcaption\label{fig:hfdist-b}
\includegraphics[width=0.8\textwidth]{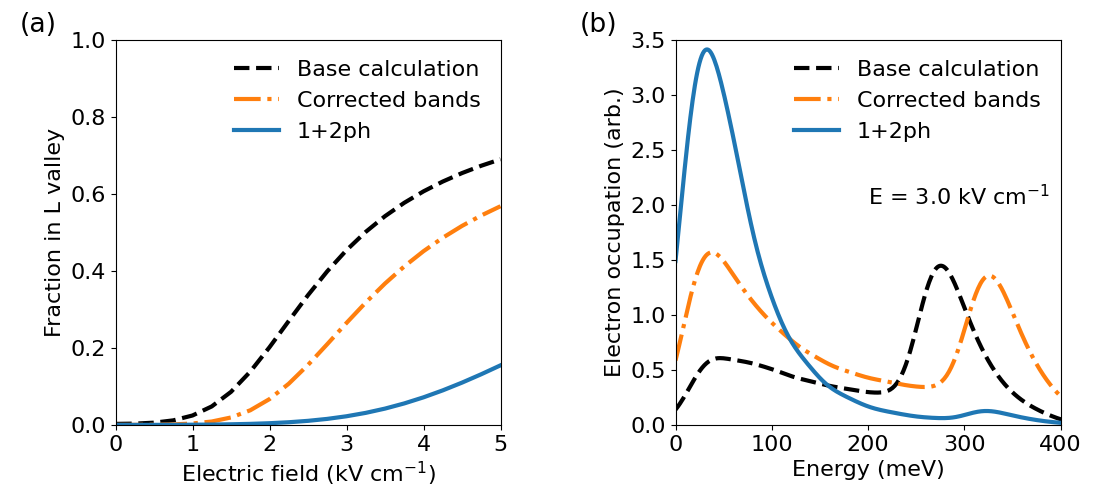}}
\caption{(a) Fraction of carriers in the L valley at steady state versus electric field for the three cases described in the text. The 1+2ph case has the least occupation  in the L valley despite having increased intervalley scattering compared to the other two cases.  (b) Steady-state distribution function versus energy at 3 \kvcm~for the three cases.}
\label{fig:hfdist}
\end{figure}

We now consider how the steady-state distribution function and valley occupations are altered by the inclusion of on-shell 2ph scattering.  Intervalley scattering causes the transfer of population from the lower-effective-mass \gam~valley to the higher-effective-mass L valley, which is the origin of negative differential resistance underlying the Gunn effect. We first investigate this transfer by plotting the fraction of the steady-state population in the L valley versus electric field in Fig.~\ref{fig:hfdist-a}. We observe that the base calculation predicts the most carriers in the L valley, followed by the corrected-bands case and then the 1+2ph case. This feature can be partly attributed to the L valley being lower in energy in the DFT bands compared to the other two cases. In the corrected-bands case where the valley separation was increased by 50 meV and the effective mass increased, fewer electrons have sufficient energy to transfer, and hence the L valley population is lower. However, the 1+2ph case has substantially fewer carriers in the L valley than even the corrected-bands case, despite having increased intervalley scattering rate as shown in Fig.~\ref{fig:2phrates-c}.

To identify the origin of this feature, we plot the steady-state distribution function for an electric field of 3 \kvcm~in Fig.~\ref{fig:hfdist-b}. The distribution function for the base calculation exhibits a clear peak around 250 meV, indicating that substantial population has transferred to the L valley. In the corrected-bands case, the peak is weaker and begins at 300 meV, reflecting the rigid shift in the L valley energy. The corrected-bands distribution function also has a higher population in the \gam~valley, consistent with Fig.~\ref{fig:hfdist-a}, due to the higher effective mass which inhibits the heating of the carriers. Finally, the distribution for the 1+2ph case exhibits still higher population for energies below 200 meV and significantly reduced L valley population.

\subsection{Process-resolved 2ph scattering rates and energy relaxation}\label{sec:eloss}

These features of the distribution function in the on-shell 2ph case reflect the increased momentum and energy dissipation contributed by 2ph processes. First, the on-shell 2ph scattering rates increase the total scattering rate by about 50\%, decreasing the mobility by around the same factor. Recall that the Joule heating per carrier is given by $e \mu E^2$ (Eq.~3.100a in Ref.~\cite{GGK_1979}). The heating of the electrons therefore decreases with the on-shell 2ph rates added. In addition to this reduced Joule heating, energies of the final states reached by 2ph scattering processes involve combinations of phonons and thus the energy relaxation mechanisms due to 2ph processes may qualitatively differ from those of 1ph processes in which phonons are only emitted or absorbed.

To gain more insight into energy relaxation by 2ph processes, we first disaggregate the 2ph scattering rate by the process type. Recall that there are three types of 2ph processes: phonon emission combined with phonon absorption (1e1a), emission of two phonons (2e), and absorption of two phonons (2a). For 1ph scattering events, the electrons gain energy when absorbing a phonon and lose energy when emitting a phonon. For 2ph scattering, 2a processes cause energy gain, 2e processes cause energy loss, and 1e1a processes lead to little energy change when mediated by phonons of similar energy, which is approximately true for 2ph scattering below 200 meV involving only LO phonons with little dispersion. When considering only LO phonon processes, the 2e and 2a processes produce approximately twice the energy loss or gain of the corresponding 1ph process. Thus, 2ph processes may substantially alter the energy relaxation compared to the 1ph case.

% Above 100 meV,  2e processes (red triangles) make the largest contribution of the three types. 

\begin{figure}[h]
    \centering{
    \phantomsubcaption\label{fig:eloss-a}
    \phantomsubcaption\label{fig:eloss-b}
    \phantomsubcaption\label{fig:eloss-c}
    \phantomsubcaption\label{fig:eloss-d}
    \includegraphics[width=0.8\textwidth]{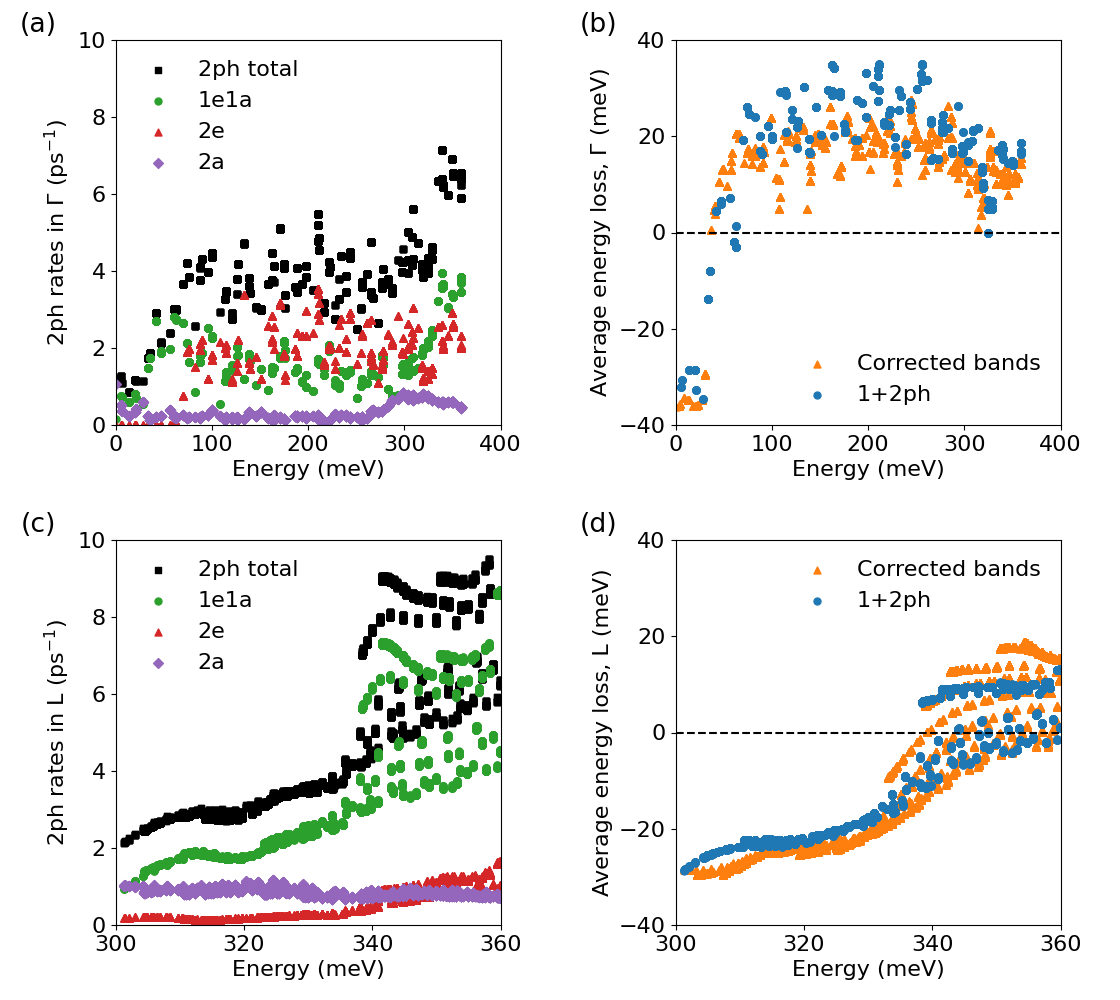}}
    \caption{2ph rates by process type and corresponding energy loss. (a) Breakdown of 2ph process type versus energy in the \gam~valley. At low energies, the 1e1a rates  comprises nearly all of the total 2ph rates. The 2a rates make a minor contribution at all energies. (b) Average energy loss versus electron energy in the \gam~valley. The 1+2ph case has  higher energy loss between 100 meV and 300 meV where the 2e rates are strongest. (c) Breakdown of 2ph process type versus energy in the L valley. The 1e1a rates are the dominant type, with 2a rates having a weak energy dependence and contributing most at the L valley minimum. (d) Average energy loss versus energy in the L valley. }
    \label{fig:eloss}
\end{figure}

% At higher energies, the difference decreases since the 2ph rate is a smaller fraction of the total scattering rate.
%  showing the less negative (closer to zero) energy loss of the 1+2ph case near the valley minimum due to the dominance of 1e1a processes

The scattering rate for each type of 2ph process in the \gam~valley is shown in Fig.~\ref{fig:eloss-a}. We observe that at low energies below 60 meV, only 1e1a and 2a processes are present, with the 1e1a being the dominant process. This result is consistent with that reported for the full 2ph calculation (Fig.~5 in Ref.~\cite{lee2020}). No 2e processes are allowed below 60 meV since the emission of two LO phonons would result in a final state with an energy in the band gap. Starting around 60 meV, 2e processes are energetically allowed, and above 100 meV they are larger in magnitude than the 1e1a processes. The 1ph case exhibits less structure because the only allowed processes are single phonon absorption and emission, leading to a single transition to phonon emission dominated scattering at around 35 meV as in Fig.~\ref{fig:2phrates-a}. 

% As intervalley transitions become possible around 275 meV, 2a and 1e1a processes increase their contribution, while 2e processes do not increase until slightly above 300 meV since reaching the L valley minimum through emission of two phonons requires starting at higher energies in the \gam~valley.
 
We next compute the average energy loss versus the energy of electrons in the \gam~valley. The average energy loss for a state at wave vector $\kwv$ is given by the following equation:

\begin{equation}
    \langle\epsilon^{\mathrm{loss}}\rangle_{\kwv} = 
    \frac{1}{\Gamma_\kwv} 
    \sum_{\kwv'}(\epsilon_{\kwv} - \epsilon_{\kwv'})
    \Theta_{\kwv',\kwv}
\end{equation}
where $\Gamma_\kwv$ is the total scattering rate for state $\kwv$ due to all scattering processes, and $\epsilon_\kwv$ is the energy of the state at $\kwv$. This weighted average quantifies the average energy exchanged with the lattice by an electron after scattering considering all types of emission and absorption processes. A positive energy loss means that, on average, carriers at that energy tend to emit phonons and lose energy, while a negative energy loss means that carriers tend to absorb phonons and gain energy. 

The result for the corrected-bands and 1+2ph cases are given in Fig.~\ref{fig:eloss-b}. Below 35 meV, the average energy loss is negative with a value around $-35$ meV for the corrected-bands case, corresponding to the LO phonon absorption dominated scattering. The 1+2ph case shows a slightly less negative energy loss because the 1e1a processes, which dominate at low energy, are nearly elastic,  shifting the average energy loss towards zero. Above 35 meV in the corrected-bands calculation, LO phonon emission processes are energetically allowed and begin to dominate the scattering, leading to a positive energy loss. For the 1+2ph case, as 2e processes start to dominate above 100 meV, the average energy loss increases and ends up about 20\% higher (5 meV) than the corrected-bands result between 100 and 250 meV. Near the energy of the L valley minimum, the increased contribution of 1e1a and 2a processes above 275 meV reduces the difference in energy loss between the 1+2ph and the corrected-bands case.

The observation of higher average energy loss for 2ph processes helps to explain the decreased population of carriers in the L valley for the 1+2ph case. In addition to the decreased power input from Joule heating owing to the lower mobility, 2ph processes are able to more effectively cool the electronic system, decreasing the population with sufficient energy to transfer to the L valley.

We now examine the categorization by 2ph scattering processes and the average energy loss for the L valley, shown in Figs.~\ref{fig:eloss-c} and \ref{fig:eloss-d}, respectively. For the scattering categorization, we observe a qualitatively similar trend as in the \gam~valley, where the majority of the 2ph scattering near the L valley minimum is 1e1a, with 2a scattering rates depending only weakly on energy. The 2e rates increase rapidly around 330 meV and exceed the 1e1a rates near the edge of the energy window (360 meV).

Next, we plot the average energy loss from the 1+2ph and corrected-bands cases in Fig.~\ref{fig:eloss-d}. The corrected-bands case with 1ph scattering has negative energy loss (energy gain) below 340 meV because scattering is dominated by phonon absorption, as was the case for the \gam~valley. Over all energies, the magnitude of the average energy loss in the 1+2ph case is  less than that of the corrected-bands case due to the dominance of the nearly elastic 1e1a processes.

% higher between 300 meV and 340 meV. Above 340 meV, the difference between the two cases is smaller due to the 2ph rates in the L valley becoming less significant in comparison to the 1ph rates, as discussed for Fig.~\ref{fig:2phrates-d}. The reason for the less negative energy loss near the L valley minimum is the same as in the \gam~valley; namely, the 1e1a processes are nearly elastic, causing the average energy loss to be less negative.

\subsection{Current PSD}

\begin{figure}[h]
\centering
    \includegraphics[width=0.6\textwidth]{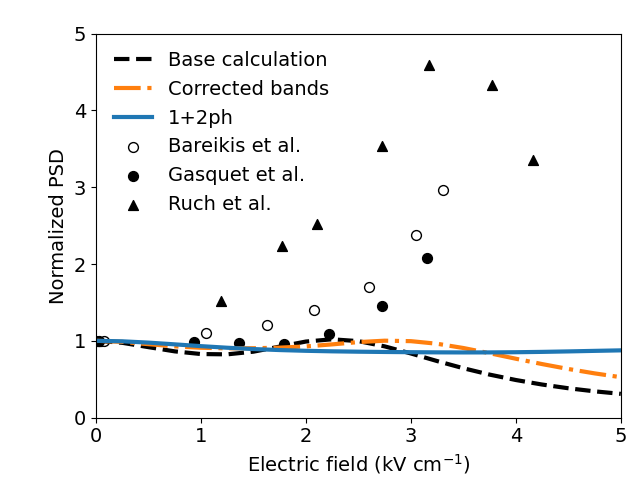}
    \caption{Normalized PSD versus electric field for the three cases as described in the text. None of the calculated cases are able to reproduce the peak in the PSD near 3 \kvcm~that appears in the experimental data: PSD from noise temperature and differential mobility measurements (filled circles \cite{gasquet1985} and open circles \cite{Bareikis_1980}), and from time of flight experiments (triangles \cite{ruchkino1968}).}
    \label{fig:psd}
\end{figure}

The PSD is sensitive to the strength of intervalley scattering processes \cite{pozela1978, Stanton_1987_2}, and hence a stricter test of the present level of theory can be obtained by computing the PSD of the hot electrons. The normalized hot electron PSD versus electric field along with experimental measurements from the literature is given in Fig.~\ref{fig:psd}. The experimental PSD exhibits a characteristic non-monotonic trend of an initial decrease, followed by a marked increase around the onset of negative differential mobility and a subsequent decrease. The data have been obtained by various methods including time of flight \cite{ruchkino1968} for the diffusion coefficient and direct measurements of noise power \cite{Bareikis_1980, gasquet1985}. The time of flight results of Ref.~\cite{ruchkino1968} are suggested to  overestimate the magnitude of the peak \cite{glisson1980}, but despite some spread in the data and the possibility of experimental inaccuracies, the same qualitative trend has been reproduced in several studies.

This trend has been attributed to the following factors. First, the PSD weakly decreases  at low fields as carriers are heated to higher energies with higher scattering rates, leading to a decrease in mobility. As a result, the PSD also decreases since the proportionality relation between mobility and PSD \cite{Kogan_1996, Callen_1952, Nyquist_1928} is approximately satisfied at low fields.  Near the threshold field for negative differential mobility, intervalley transitions become possible, and the peak in the PSD has been attributed to intervalley diffusion that arises due to scattering between two valleys of highly dissimilar effective masses \cite{Shockley_1966, Price_1960, Jacoboni2010}. Finally, the PSD decreases at high fields due to the accumulation of carriers in the L valley. The decrease occurs because the group velocities in the L valley are substantially lower than those in the \gam~valley and hence make a lesser contribution to the electric current and PSD as indicated by the group velocity factor in the sum in Eq.~\ref{noisesum}.

The PSD from the base calculation predicts some features of the experimental non-monotonic trend, with the initial decrease at low field originating from the increase in scattering rates \cite{warmelectrons}, followed by a weak peak and decrease above 2.5 \kvcm. However, overall the calculated PSD is in poor agreement with the experiments, with the rise in the PSD being significantly underestimated. In the corrected-bands case, the initial decrease of the PSD versus field is weaker, consistent with the increased effective mass which inhibits electron heating, and the subsequent decrease of the PSD after the peak occurs at a larger field (3 \kvcm), consistent with the L valley minimum having been shifted to higher energies. However, little improvement in the magnitude of the PSD peak is observed. Finally, the 1+2ph case yields similarly poor agreement,  with the weak peak largely unchanged compared to either calculation at the 1ph level of theory.

% Further, in all three cases the onset of intervalley scattering by itself does not produce a peak in the PSD, suggesting that the precise details of how intervalley scattering mediates transitions between valleys are of high importance to producing this characteristic feature. Further discussion on this point is given in Sec. \ref{missingpeak}.

%%% DISCUSSION
\section{Discussion}\label{disc}

We have established that the on-shell 2ph level of theory makes a non-negligible contribution to high-field transport properties but does not predict the trend of PSD in GaAs. We have also shown how the on-shell 2ph contributes substantially to intervalley scattering and qualitatively affects the evolution of the electron distribution function with electric field, in particular by increasing the energy relaxation rate. We now discuss how these findings allow for the resolution of a discrepancy in the IDP inferred from different experiments \cite{mickevicius_1990}, and we discuss the possible reasons for the lack of a peak in the computed PSD.

\subsection{Interpretation of experimental studies of intervalley scattering}

Intervalley scattering in GaAs has been the subject of intensive experimental and theoretical study owing to its importance to negative differential resistance and the Gunn effect. Many studies aimed to  quantify the strength of intervalley scattering as measured by the IDP in the semi-empirical expression originally derived by Conwell \cite{conwell_1967}. While this model is now known to be inaccurate  \cite{Sjakste2007, zollner1990_JAP}, the IDP in the model  nevertheless captures the magnitude of intervalley processes in a single number that is comparable across studies \cite{zollner1990_solst}. 

The IDP value in GaAs has been inferred primarily from two classes of experiments, charge transport and photoluminescence response to optical excitation. In transport studies, an external field was applied to a sample and the current or noise response was measured. The transport was simultaneously modeled with Monte Carlo methods based on semi-empirical scattering rates, and the IDP was obtained by fitting  simulation and experiment. This approach has been used on samples subjected to uniaxial stress along the [111] crystal axis \cite{Adams1977} to identify the shift in threshold field for onset of negative differential mobility with stress, with additional modeling performed in Ref.~\cite{mitskyavichyus1986}; and measurements of the diffusion coefficient \cite{ruchkino1968} with modeling in Ref.~\cite{pozela1978}. The value of the IDP extracted from these experiments is approximately $D \sim 2 \times 10^8$ eV cm\textsuperscript{-1}.

\begin{figure}[h]
    \centering
    \includegraphics[width=0.6\textwidth]{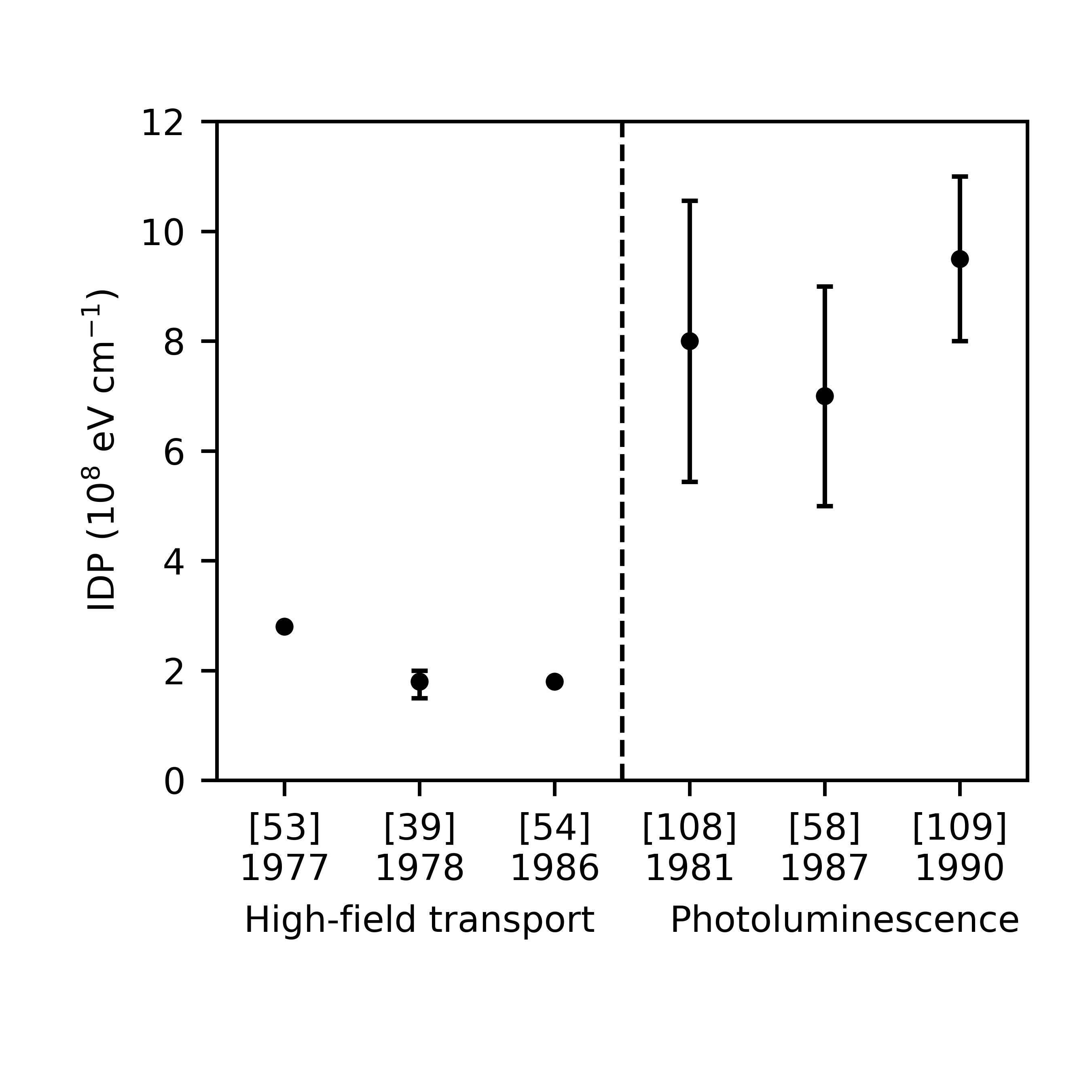}
    \caption{ Intervalley deformation potential (IDP) values inferred from high-field transport and photoluminescence experiments. The high-field transport experiments require modeling to fit the data \cite{Adams1977, pozela1978, mitskyavichyus1986}. The photoluminescence experiments extract a lifetime that is used to deduce the IDP \cite{dymnikov1981, karlik1987, fasol1990}. A clear discrepancy exists in the IDP inferred from the two sets of experiments.}
    \label{fig:idp}
\end{figure}

In the other class of experiments, the sample was subjected to optical excitation and the resulting photoluminescence was measured. Although the details vary between experiments, the electron lifetime below and above the L valley energy can be directly extracted from the measurements, thereby providing the intervalley scattering rate. This approach does not require assumptions regarding the physical origin of the scattering. This approach has been employed by Dymnikov et al., who measured the depolarization of photoluminescence in a magnetic field \cite{dymnikov1981}; Karlik et al., who deduced a lifetime based on relative photoluminescence intensities \cite{karlik1987}; and Fasol et al. using the broadening of the photoluminescence peak \cite{fasol1990}. The value of the IDP from these methods is generally around a factor of four larger than that inferred from transport studies. We show the differing values of IDP from both classes of experiments in Fig.~\ref{fig:idp}. An extensive review of the discrepancy was given by Reklaitis et al. \cite{mickevicius_1990}, and it remains unresolved.

Our observation that on-shell 2ph scattering has a significant effect on the high-field transport properties in GaAs provides a means to reconcile the differing conclusions. An important difference between the two types of experiments is that determining the intervalley scattering strength from transport experiments requires interpretation using  simulations, while the optical experiments directly provide a lifetime. The IDP values inferred from transport experiments are therefore susceptible to inaccuracies in the assumed scattering rates. We have shown in Fig.~\ref{fig:hfdist-b} that the inclusion of 2ph scattering qualitatively changes the steady-state distribution function at high fields due to contributions to momentum as well as energy relaxation (Figs.~\ref{fig:eloss-b} and \ref{fig:eloss-d}). In the MC studies used to interpret transport experiments, semi-empirical 1ph rates were adjusted to predict the experimental low-field mobility. However, these scattering rates underpredict the energy relaxation and hence overpredict the population at high energies, leading to an excess population in the L valley and hence a suppression of the PSD peak. A lesser value of the IDP would therefore be needed to compensate, explaining the smaller value inferred from transport studies. Conversely, if the semi-empirical 1ph rates were adjusted to prevent the overpopulation at high energies, the low-field mobility would be substantially underpredicted. The coupling between momentum and energy relaxation at the 1ph level of theory is altered by the inclusion of 2ph scattering.

We therefore conclude that the correct intervalley scattering rates  are those inferred from optical studies, resolving the discrepancy regarding the strength of intervalley scattering in GaAs as described in Ref.~\cite{mickevicius_1990}. Our work also provides a clear physical origin for the underprediction of the IDP from transport studies.

\subsection{Possible origin for lack of PSD peak}\label{missingpeak}

Finally, we consider candidate origins of the discrepancy in the PSD versus electric field. One possibility is that off-shell processes that are neglected in the present calculation are necessary. However, as in Fig.~\ref{fig:2phrates-a}, the computed on-shell 2ph rates agree well with those from the full calculation in the range of energies for which comparison is possible, implying that the present rates have captured most of the relevant processes.  Another mechanism may be a non-trivial cancellation of the next-leading-order term of electron-phonon interaction with the  term involving the second derivative of the interatomic potential. \cite{kocevar_1980} This cancellation has long complicated the investigation of higher-order electron-phonon interactions. Investigating these hypotheses with ab-initio methods is computationally challenging but an interesting target of future study.

Finally, a more prosaic explanation could be experimental non-idealities. From the original study of the Gunn effect, it was observed that current noise occurred owing to the instability associated with  negative differential resistance, which in turn arose from the formation of charged domains. \cite{gunn1964} In general, ab-initio simulations do not consider space charge effects, and so the present simulations would not predict any effect arising from the dipole layers. An inconsistency with this explanation is the appearance of the peak in PSD prior to the electric field at which NDR occurs. However, this inconsistency might be accounted for by imperfections in electrical contacts and doping fluctuations leading to local electric fields that exceed the NDR threshold \cite{mccumber_1966}. The PSD discrepancy is a topic of future study.

\section{Conclusions}

We have reported the high-field drift velocity and hot-electron PSD in GaAs from first principles. Although the 1ph theory has been thought to be adequate for GaAs and used extensively in Monte Carlo simulations, we have found that on-shell 2ph processes play a fundamental role in all aspects of high-field transport, including energy relaxation and intervalley scattering. This finding resolves a long-standing discrepancy regarding the value of the IDP as inferred from transport and optical studies in favor of the stronger value obtained from photoluminescence measurements. Further, the characteristic peak in the PSD versus electric field is not predicted at this level of theory. Our work demonstrates that the \textit{ab initio} computation of high-field transport and noise properties may provide considerable insight into the e-ph interaction in semiconductors.

\section{Acknowledgements}

This work was supported by AFOSR under Grant Number FA9550-19-1-0321. The authors thank B.~Hatanp\"a\"a, D.~Catherall and T.~Esho for helpful discussions.

\bibliography{references}
 
\end{document}